\documentclass{aa}
\usepackage{graphicx}
\begin{document}
   \title{Jet/cloud collision, 3D gasdynamic simulations of HH 110}

   \author{A. C. Raga\inst{1}, E. M. de Gouveia Dal Pino\inst{2},
A. Noriega-Crespo\inst{3}, P. D. Mininni\inst{4}, P. F. Vel\'azquez\inst{1}}

   \offprints{A. Raga}

   \institute{Instituto de Ciencias Nucleares, UNAM, Ap. 70-,
04510 D. F., M\'exico\\
             \email{raga@astroscu.unam.mx,pablo@nuclecu.unam.mx}
   \and
   Instituto de Astronomia, Geofisica e Ciencias Atmosfericas,
Universidade de S\~ ao Paulo, R. do Mat\~ ao 1226, 055-08-090 S\~ ao
Paulo, SP, Brasil\\
             \email{dalpino@astro.iag.usp.br}
   \and
   SIRTF Science Center,
California Institute of Technology, IPAC 100-22, Pasadena,
CA 91125, USA\\
             \email{alberto@ipac.caltech.edu}
   \and
   Departamento de F\'\i sica, Facultad de Ciencias Exactas
y Naturales, Universidad de Buenos Aires, Ciudad Universitaria,
1428 Buenos Aires, Argentina\\
             \email{mininni@df.uba.ar}
             }

   \date{}

   \abstract{We present 3D, gasdynamic simulations of jet/cloud collisions,
with the purpose of modelling the HH~270/110 system. From the models,
we obtain predictions of H$\alpha$ and H$_2$ 1-0 s(1) emission line maps,
which qualitatively reproduce some of the main features of the corresponding
observations of HH~110. We find that the model that better reproduces
the observed structures corresponds to a jet that was deflected at
the surface of the cloud $\sim 1000$~yr ago, but is now boring a tunnel
directly into the cloud. This model removes the apparent contradiction
between the jet/cloud collision model and the lack of detection of
molecular emission in the crossing region of the HH~270 and HH~110
axes.
   }

\authorrunning{Raga et al.}
\titlerunning{Jet/cloud collision simulations}

   \maketitle

\section{Introduction}

The Herbig-Haro (HH) jet HH~110 (discovered by Reipurth \& Bally 1986)
is the best observed example of a possible HH jet/dense cloud collision.
Reipurth et al. (1996) have interpreted the rather unique, collimated
but quite chaotic structure of HH~110 as the result of a deflection of
the faint HH~270 jet through a collision with a dense cloud.

%                                     Two column figure
   \begin{figure*}
   \centering
   \includegraphics{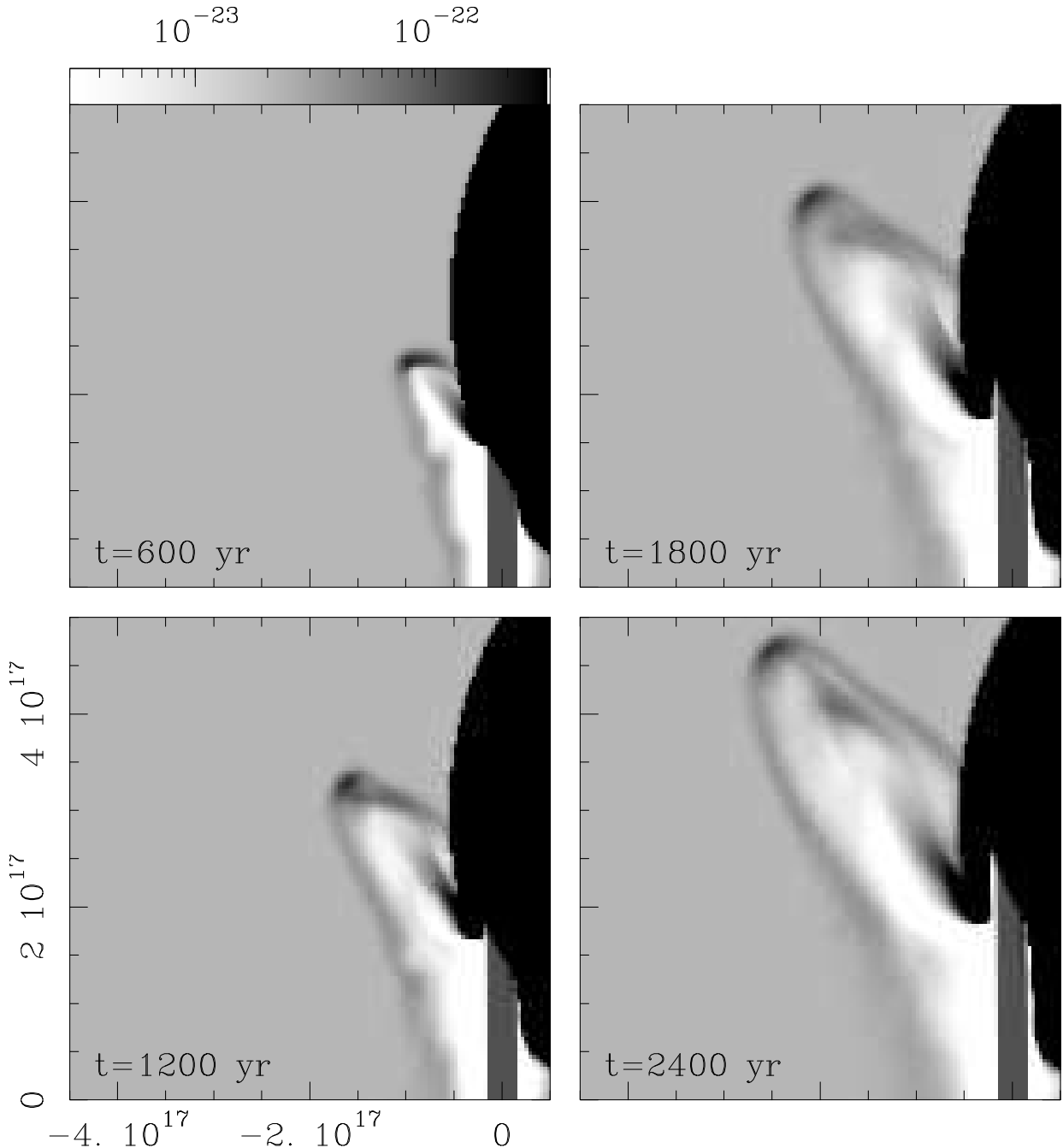}
   \caption{Time sequence of the density stratifications obtained
from model A. The density stratifications on the $y=0$ plane
(which includes the outflow axis and the centre of the spherical
cloud) are shown for different integration times (as indicated
at the bottom left of each plot). The densities are depicted
with a logarithmic greyscale, with the values given (in g~cm$^{-3}$)
by the bar on the top left of the figure. The $x$ (horizontal)
and $z$ (vertical) axes are labeled in cm.}
              \label{dens}
    \end{figure*}

The evidence presented by Reipurth et al. (1996) for this interpretation
can be summarized as follows:
\begin{itemize}
\item no stellar source has been detected aligned with the HH~110 jet,
\item the HH~270 jet (ejected from a detected IR and radio source,
see Rodr\'\i guez et al. 1998) points towards
the beginning of the HH~110 jet,
\item the proper motions of HH~270 and HH~110 have an approximately
2 to 1 ratio, which is completely consistent with the $\approx 60^\circ$
deflection angle defined by the locci of the two jets (the flow
approximately lying on the plane of the sky).
\end{itemize}
This last statement can be understood as follows. When a radiative
jet hits the surface of a dense cloud at an incidence angle $\phi$
(between the incident jet axis and the cloud surface), the normal
component of the jet velocity is stopped in a radiative shock,
and the jet continues to flow parallel to the surface (conserving
the component of the incident jet velocity parallel to the surface).
Therefore, the velocity $v_{def}$ of the deflected jet is approximately
equal to the projection of the incident jet velocity $v_{inc}$ along
the surface of the dense cloud. Therefore, $v_{def}\approx v_{inc}\cos\phi$.
One can straightforwardly see that the proper motions and deflection
angle defined by HH~270 and HH~110 (see above) do satisfy this condition.

This result led Raga \& Cant\'o (1995) to study the dynamics of the
collision of a radiative, HH jet with the surface of a dense cloud.
These authors presented an analytic model and plane, 2D simulations
of the early stages of such an interaction, and found that the general
characteristics of the HH~270/110 system could be explained in terms
of such a model. The main problem found with the models is that
in a rather short timescale, the incident jet starts to perforate
the obstacle, and the deflected jet beam is then pinched off. In order
to obtain a long enough timescale for the production of the
deflected jet, it is necessary to have a very high
cloud-to-jet density ratio. Raga \& Cant\'o (1995) suggested that
this problem might be overcome if the incident jet did not have
a completely straight jet beam, so that the impact point would roam
over the surface of the dense cloud.

%                                                One column figure
   \begin{figure}
   \centering
   \includegraphics[width=8cm]{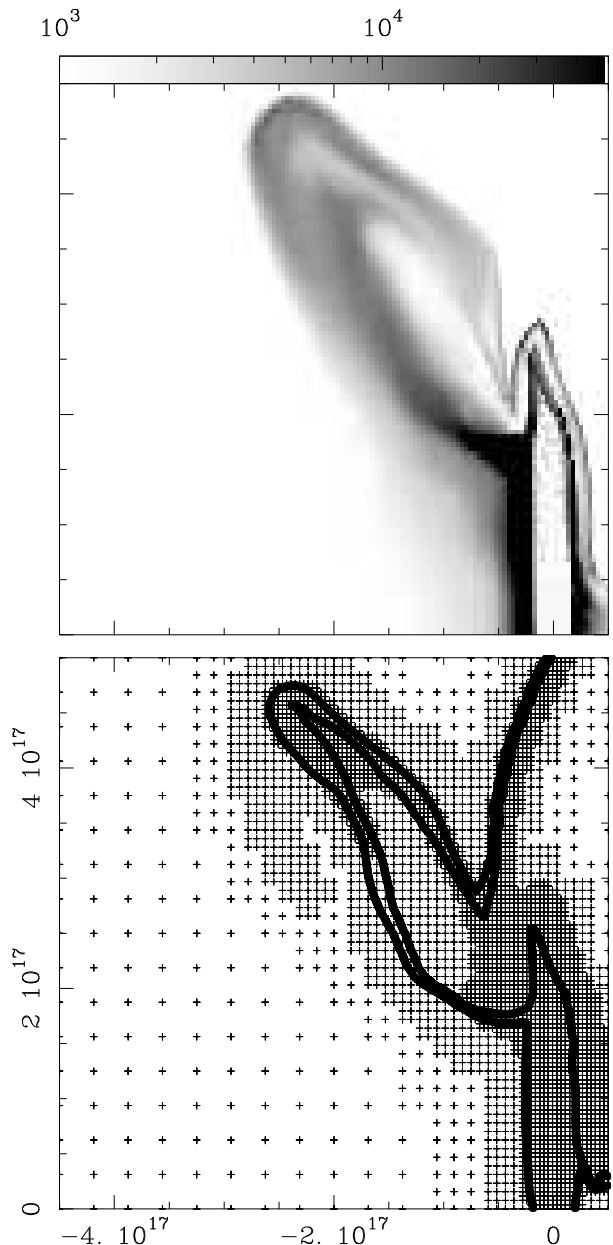}
      \caption{Temperature stratification (top) and adaptive grid
structure (bottom) on the $y=0$ plane obtained from model A
for a $t=2600$~yr integration time. The temperature stratification
is depicted with a logarithmic greyscale with the values given
(in K) by the bar on the top of the figure. In the bottom plot,
two thick lines separating the jet, cloud and environmental
material are shown (see the text). These lines show two values
of the passive scalar~: $\psi=0$ (outer contour) and $\psi=1.5$
(inner contour).
              }
         \label{temp}
   \end{figure}

The regime in which the jet has punched a hole through a cloud
was described by Cant\'o \& Raga (1996) and Raga \& Cant\'o (1996).
If the cloud is stratified, the path of the jet through the cloud
is curved, though the curvature is important only if the radius
of the cloud is comparable to the jet radius. 3D gasdynamic simulations
of the penetration of a jet into and through a dense, stratified cloud
were carried out by de Gouveia Dal Pino (1999).

Finally, Hurka et al. (1999) have studied the bending of the beam of
a 3D MHD, non-radiative jet by a magnetic field with a strong gradient
(as would be found at the surface of a dense cloud). These authors
show that this effect would help to increase the timescale over
which the jet/cloud surface interaction takes place, before the
deflected jet is pinched off.

In the present paper, we discuss 3D gasdynamic simulations of
the interaction of a radiative jet with the surface of a dense
cloud. We show the results from two simulations with different
assumptions for the incident jet~:
\begin{itemize}
\item that the jet is ejected with a constant direction and velocity
\item that it is produced with a precessing outflow direction
and a sinusoidally varying velocity.
\end{itemize}
Through a comparison of these two simulations, we can evaluate the
effect of a ``roving'' impact point on the production of the
deflected jet.

%                                     Two column figure
   \begin{figure}
   \centering
   \includegraphics[width=8cm]{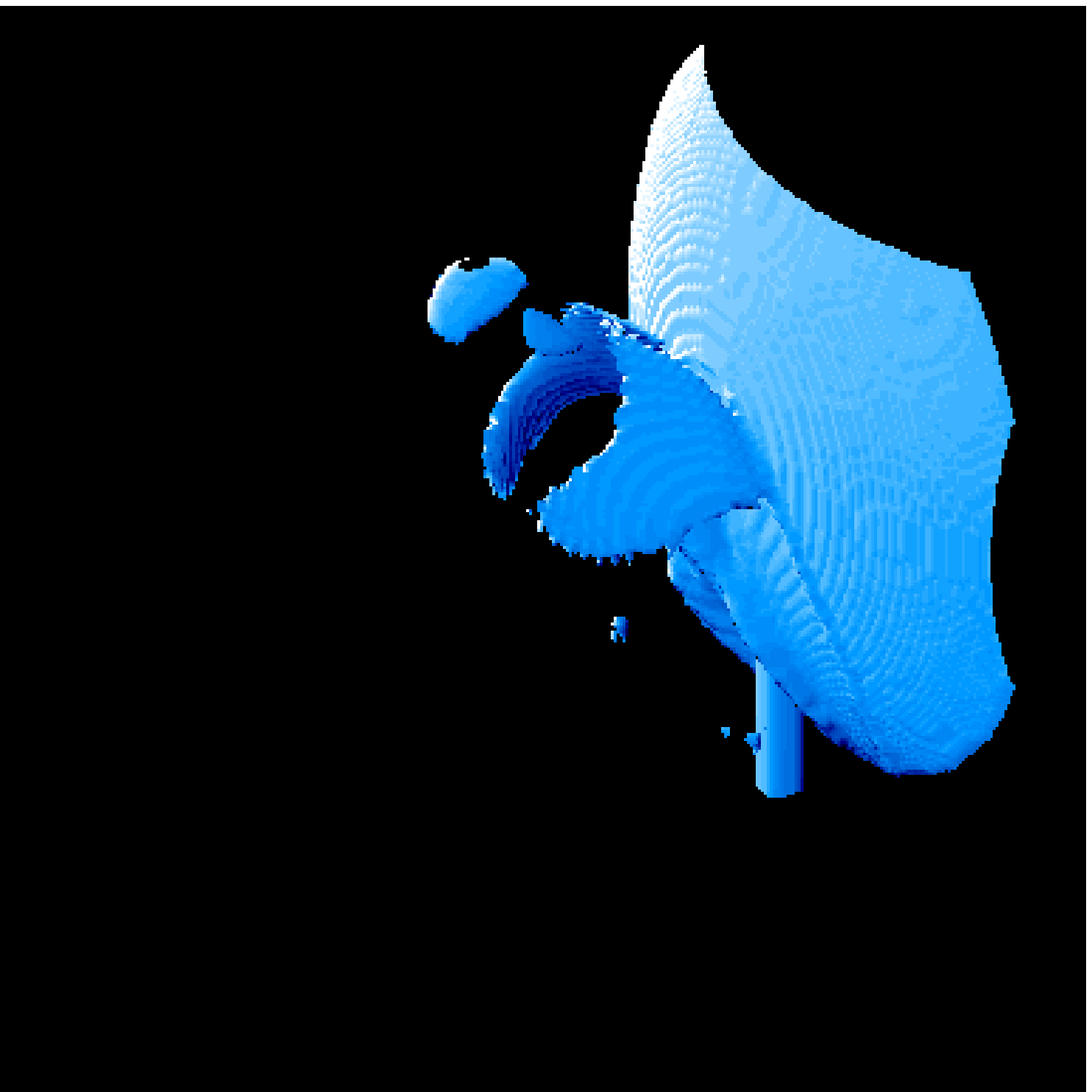}
   \includegraphics[width=8cm]{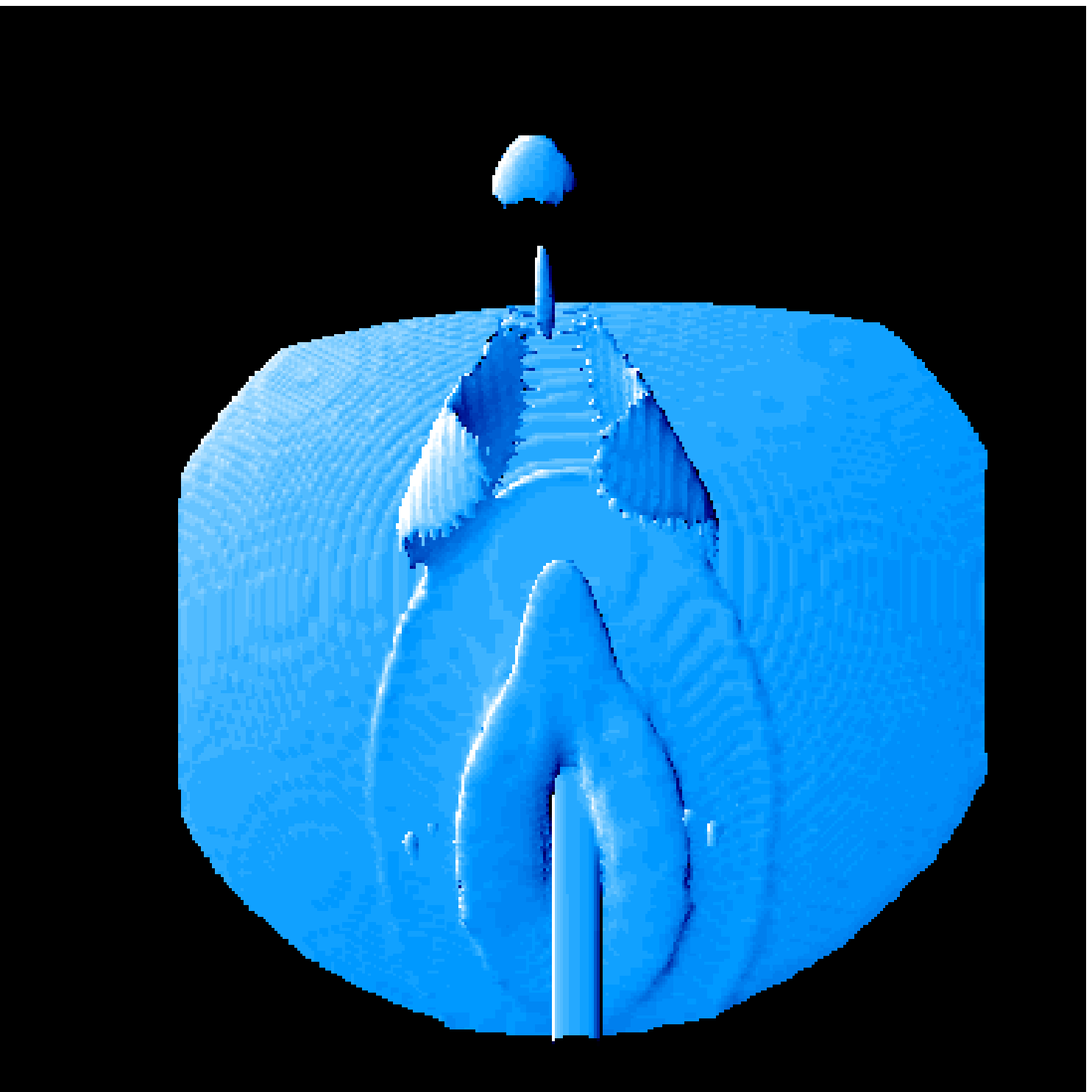}
   \caption{Constant density surface (corresponding to a
$n=20$~cm$^{-3}$ number density of atomic nuclei) from model A
for a $t=2600$~yr integration time. The two graphs show the surface
as seen from two different directions.}
              \label{3d}
    \end{figure}

Our simulations include a treatment of the dissociation and ionization
of the gas. Therefore, we can use the results to obtain predictions
of atomic and molecular lines, which we directly compare with
previously published images of HH~110.

In particular, we compute predicted maps in the H$_2$ 1-0 s(1) line.
This is of interest because the morphology of HH~110 in this IR
line is quite strikingly different from its morphology in
atomic/ionic lines. Davis et al. (1994) and
Noriega-Crespo et al. (1996) found
that the H$_2$ emission is much better collimated, and lies along
one of the edges of the HH~110 jet beam. This led Noriega-Crespo
et al. (1996) to present a simple model of the molecular emission
as coming from material from the dense cloud which is entrained
by the jet as it brushes past the cloud surface. Our present
simulations allow us to make a more definite assesment of whether
or not such a mechanism actually succeeds in explaining the molecular
emission of HH~110.

We should point out that Choi (2001) present HCO$^+$ emission
maps, in which they detect emission in the HH~270/110 region, but not
around the ``point of impact'' in which the ``incident'' HH~270 jet
is presumably redirected into the ``deflected'' HH~110 jet. This
result leads them to suggest that HH~110 might actually not be
the result of a jet/cloud collision, but that it could instead
be a ``straight'' jet ejected from a low luminosity, undetected
stellar source which is presumably more or less aligned with
the direction of the HH~110 flow. In the conlcusions, we discuss
the possible ways of reconciling the jet/cloud interaction model
with the observations of Choi (2001) which are suggested
by our 3D gasdynamic simulations.

\section{The numerical simulations}

\subsection{General features}

We have carried out 3D gasdynamic simulations of jet/cloud interactions
with the yguaz\'u-a adaptive grid code. This code integrates the
3D (or 2D) gasdynamic equations together with a set of continuity
equations for atomic/ionic or chemical species. The details of the
gasdynamic and the adaptive grid algorithms have been presented
by Raga et al. (2000), and tests of the code are given by Sobral et al.
(2000) and Raga et al. (2001).

For the present simulations, we have used the following set
of species~: H$_2$, H~I and II, C~II, III and IV, and O~I, II and III
(with abundances by number relative to hydrogen~:
$y_C={6.6\times 10^{-4}}$ and $y_O={3.3\times 10^{-4}}$).
For the atomic/ionic reactions, we have included the collisional
ionization (from Cox 1970), radiative+dielectronic recombination
(from Aldrovandi \& P\'equignot 1973, 1976) and O/H charge exchange
processes. The H$_2$ dissociation and cooling has been included
in the same way as in Raga et al. (1995). The cooling associated
with the atomic/ionic species has been included as described
in appendix A.

We have computed two jet/cloud interaction models, which share
the following characteristics. In both models, an initially
atomic jet (except for C, which is singly ionized) of
number density $n_j=50$~cm$^{-3}$ and temperature $T_j=1000$~K
interacts with a spherical, homogeneous molecular cloud
(with all H in the form of H$_2$) of number density
$n_c=5000$~cm$^{-3}$ and $T_c=1$~K. The cloud is surrounded
by a homogeneous, neutral environment of density $d_{env}=10$~cm$^{-3}$
and temperature $T_{env}=1000$~K. Therefore, the jet to cloud
(mass) density ratio is $\rho_j/\rho_c=1/100$.

%                                     Two column figure
   \begin{figure*}
   \centering
   \includegraphics{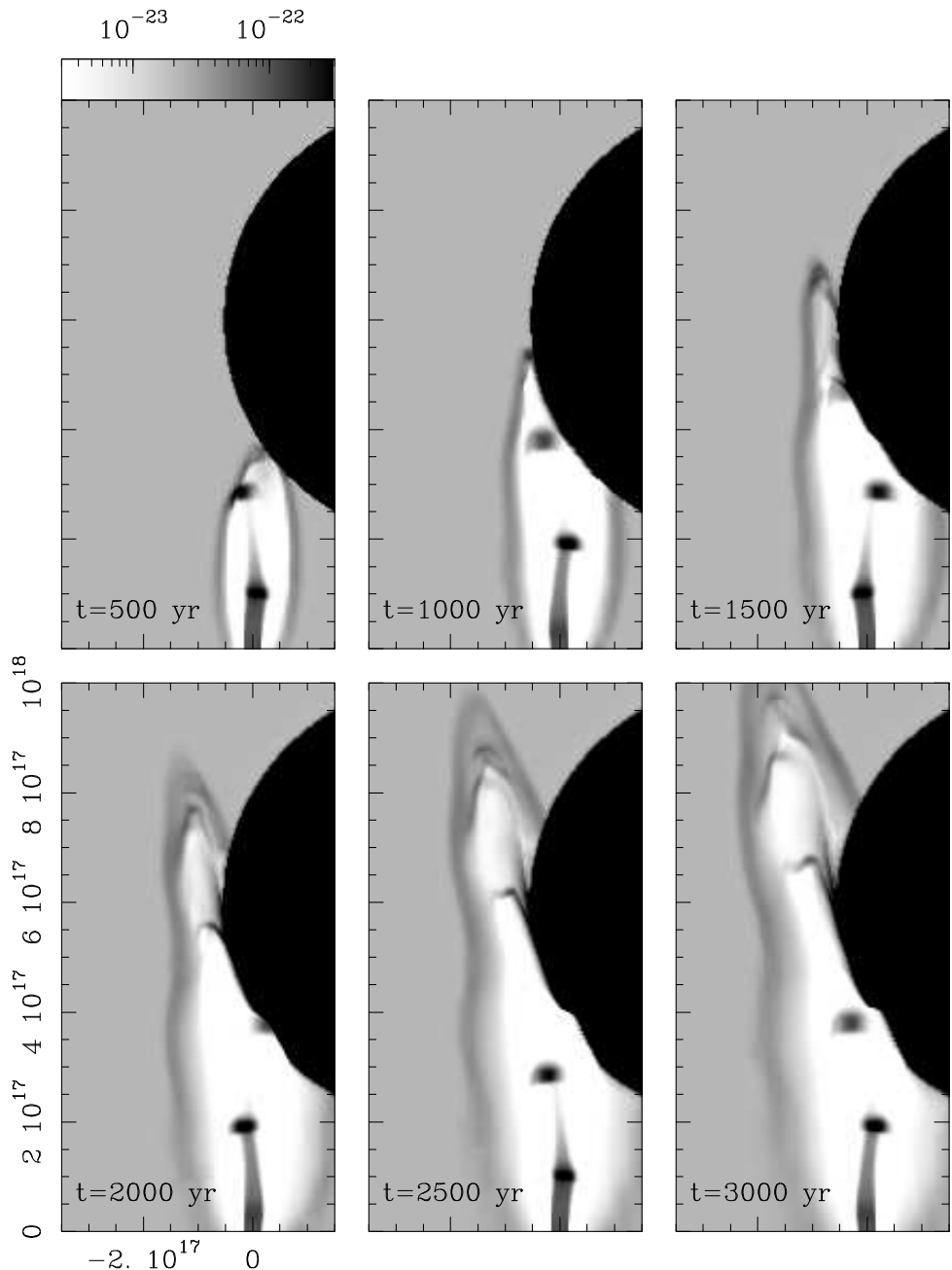}
   \caption{Time sequence of the density stratifications obtained
from model B. The density stratifications on the $y=0$ plane
(which includes the outflow axis and the centre of the spherical
cloud) are shown for different integration times (as indicated
at the bottom left of each plot). The densities are depicted
with a logarithmic greyscale, with the values given (in g~cm$^{-3}$)
by the bar on the top left of the figure. The $x$ (horizontal)
and $z$ (vertical) axes are labeled in cm.}
              \label{densb}
    \end{figure*}

In both simulations, the cloud has a $r_c={4\times 10^{17}}$~cm
radius, and the jet has an initial, top-hat cross section of radius
$r_j={1.5\times 10^{16}}$~cm. The jet is injected at $(x,y,z)=
(0,0,0)$ with the outflow axis in the $z$-direction. Free
outflow conditions are applied on all of the outer boundaries
of the computational domain, except for the $z=0$ plane,
on which a reflection condition is imposed outside of
the injection jet cross section.

The computations are carried out on a 5-level, binary adaptive
grid (the two coarsest levels being defined over the full
computational domain, see Raga et al. 2000) with a maximum
resolution (along the three axes) of $3.91\times 10^{15}$~cm.
The highest resolution grid is only allowed in the regions
occupied either by jet or by cloud material (which are traced
by advecting a passive scalar), so that the maximum resolution
allowed in the environmental material is only of
$7.81\times 10^{15}$~cm.

We have then computed two models, one with a jet with time-independent
injection conditions (model A), and one with a precessing, variable
ejection velocity jet (model B). These two models are discussed
in the following two subsections.

\subsection{Jet with time-independent injection (model A)}

In this simulation, the jet is injected parallel to the $z$-axis,
with a constant $v_j$=300~km~s$^{-1}$ injection velocity. The
computational domain extends from ${-4.5\times 10^{17}}<x<
{0.5\times 10^{17}}$~cm, ${-2.5\times 10^{17}}<y<
{2.5\times 10^{17}}$~cm and $0<z<{5\times 10^{17}}$~cm. The centre
of the spherical cloud is placed at $(x_c,y_c,z_c)=
{(3.5,0,3)\times 10^{17}}$~cm, so that the jet has a glancing
collision with the surface of the cloud.

Figure 1 shows a time series (spanning an integration time
of $t=2400$~yr) of the density stratifications on the $y=0$ plane
(this plane includes the outflow axis and the centre of the
dense cloud). In this time series one sees the incident
jet beam (injected at the origin, and travelling along the $z$-axis)
impinging on the surface of the dense cloud, and being deflected
onto a direction towards the top left of the $xz$-cuts. At
$t=1200$~yr, the jet has already dug a hole into the cloud
(this hole becoming deeper at later integration times), and
the deflected jet beam basically becomes cut off at its base.
At this time, the jet/cloud impact point lies within,
rather than at the surface of the cloud.
However, the material deflected by the cloud surface at
earlier times continues to travel away from the cloud, leaving
a complex ``wake'', joining it to the point at which the
incident jet impacted the cloud surface.

In order to illustrate the configurations adopted by the
adaptive grid, Figure 2 shows the temperature stratification
and the adaptive grid structure on the $y=0$ plane obtained
for $t=2400$~yr. It is clear from this figure that the
higher resolution is not allowed on the regions occupied
by environmental gas, so that the leading bow shock is
only resolved at the second highest resolution level.

Figure 2 also shows the following. We have integrated an advection
equation for a passive scalar $\psi$. This scalar has been given a value of
$\psi=1$ for the jet, 2 for the dense cloud, and $-1$ for the surrounding
environment. In the plot showing the adaptive grid, we have also
drawn two contours on the stratification of the passive scalar
corresponding to values $\psi=0$ (outer contour)
and $\psi=1.5$ (inner contour). The region in between the
two contours corresponds to the jet material (which has $\psi=1$).
From Figure 2 it is then clear that the jet material occupies the
injection region and the hole in the cloud, as well as a ``plug''
of material (at $(x,y)\approx {(-2.5,4.5)\times 10^{17}}$~cm)
with wings which extend towards the jet/cloud impact
point. The region in between the wings is filled in
by a tongue of cloud material which has been swept into the
deflected jet flow. We find that this entrained cloud material
has interesting observational properties, which are described in
section 3.

The structure of this dense tongue is more clearly seen in Figure~3,
which shows a constant density 3D surface (corresponding to a
$n=20$~cm$^{-3}$ number density of atomic nuclei) obtained for
a $t=2600$~yr integration time. This figure
shows the jet penetrating into the cloud, part of the bow shock,
the denser region of the deflected jet beam and the entrained
molecular gas material (which forms a structure which surrounds
the incident jet, and has an elongation in the direction
of the deflected jet beam).

\subsection{Precessing, variable ejection velocity jet (model B)}

We have computed a second simulation, in which the ejection direction
precessses around the jet axis. The precession cone has an
$\alpha=5^\circ$ half-opening angle, and a $\tau_p=400$~yr
precession period. Also, the jet is injected with a constant
density ($n_j=50$~cm$^{-3}$, see \S 2.1), but with a sinusoidally
varying ejection velocity $v_j(t)=\left(300+80\sin 2\pi
t/\tau_v\right)$~km~s$^{-1}$, with a $\tau_v=200$~yr period.

For this simulation, we choose a computational domain with the same
extent as the one of model A along the $x$ and $y$ axis
(but with a diferent centering~:
${-3.5\times 10^{17}}<x<{1.5\times 10^{17}}$~cm and
${-2.5\times 10^{17}}<y<{2.5\times 10^{17}}$~cm) but
with a larger, $0<z<10^{18}$~cm extent along the axis. The center
of the spherical cloud is placed at $(x_c,y_c,z_c)=
{(3.5,0,6)\times 10^{17}}$~cm. We have chosen to have a larger
distance from the point of injection to the jet/cloud collision
region in order to allow the internal working surfaces of
the jet (which result from the ejection velocity time-variability)
to form before colliding with the dense cloud.

   \begin{figure}
   \centering
   \includegraphics[width=9cm]{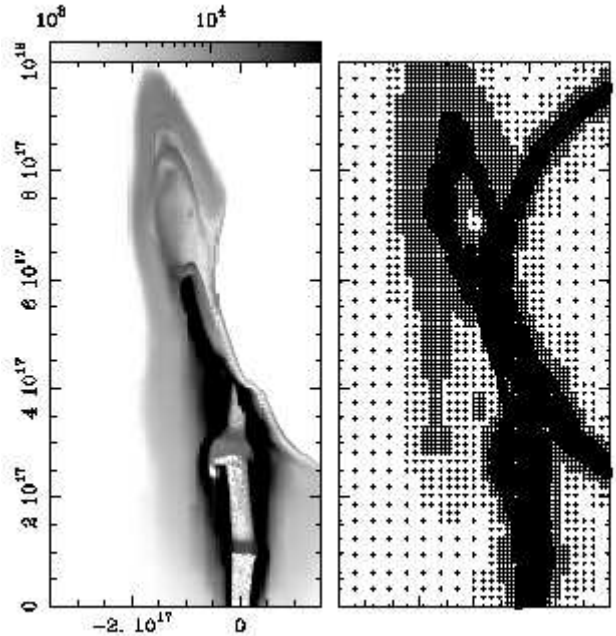}
      \caption{Temperature stratification (left) and adaptive grid
structure (right) on the $y=0$ plane obtained from model A
for a $t=2500$~yr integration time. The temperature stratification
is depicted with a logarithmic greyscale with the values given
(in K) by the bar on the top of the figure. In the right hand side
plot, two thick lines separating the jet, cloud and environmental
material are shown (see the text).
              }
         \label{tempb}
   \end{figure}

Figure 4 shows a time sequence of the density stratifications
obtained on the $y=0$ plane (which includes the precession axis
and the centre of the cloud). In this figure, one sees the internal
working surfaces which first form, and then impact the surface of the
dense cloud. Because of the precession in the ejection direction,
the successive working surfaces impact the cloud at different points.
Through a comparison with Figure 1, it is clear that this effect
increases the time that the jet takes to dig a hole into the cloud
(and therefore pinching off the deflected jet beam).
In fact, for the $t=3000$~yr time-integration shown in Figure 4,
the depth of the hole is still smaller than the diameter of
the impinging jet (for a similar time frame obtained from model A,
the depth of the hole is of approximately two jet diameters, see
Figure 1). Similar results are deduced by analyzing
different $y=const.$ cuts through the 3D density stratification.

Figure 5 shows the temperature stratification, and the grid
structure on the $y=0$ plane obtained for a $t=2500$~yr time-integration
(equivalent results for model A are shown in Figure 2). On the
graph with the adaptive grid, we again show the contours that separate
the jet, cloud and environmental material (for model A, see Figure 2
and the discussion at the end of \S 2.2). It is clear that model B
has a more complex structure than model A, showing a number of
condensations of jet material in the ``deflected flow'' region, which
correspond to the different working surfaces that have been deflected
on collision with the cloud surface.
Interestingly, the center of the deflected flow region is filled
with material from the molecular cloud, as is also the case for
model A (see Figure 2).

In the following section, we present predictions of emission line
intensity maps carried out from the results of models A and B. These
predictions can then be compared directly with the available observations
of the HH~270/110 system.

\section{Predicted intensity maps}

\subsection{General considerations}

In order to compare the jet/cloud interaction models described
in \S 2 with the published images of the HH~270/110 system, we have
obtained predicted emission line maps from the models. From the
computed temperature, density, electron density and H ionization
and molecular fractions, we have computed the H$\alpha$ and
H$_2$ 1-0 s(1) (2.12$\mu$m) emission coefficients.

%                                                One column figure
   \begin{figure}
   \centering
   \includegraphics[width=7cm]{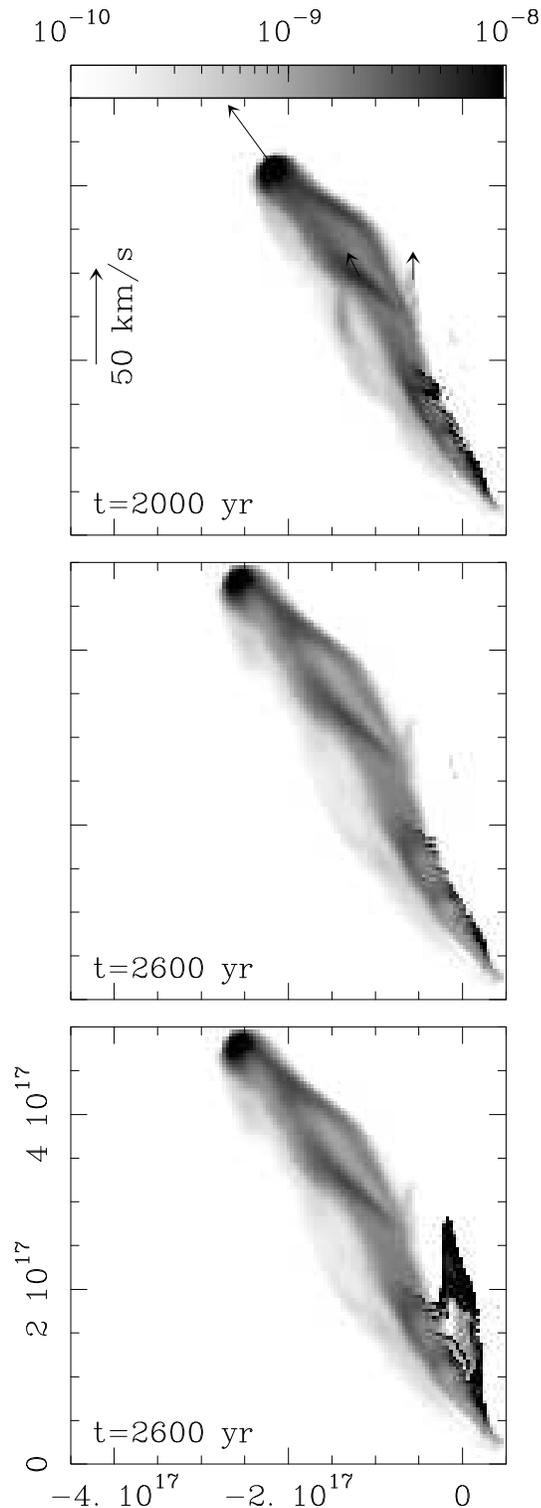}
      \caption{H$\alpha$ maps predicted from model A, corresponding
to the integration times given on the bottom left of each frame.
The maps are depicted with the greyscale given by the bar at
the top of the graph (which gives the intensity values in
erg~cm$^{-2}$~s$^{-1}$~sterad$^{-1}$).
The two top frames have been computed including the dust extinction
of the dense cloud (see \S 3.2), and the bottom frame shows a map
computed without considering this extinction. The proper motions
computed from the positions of three intensity maxima (as measured
in the $t=2000$ and 2600~yr maps) are shown in the top frame.
The $x$ (horizontal) and $z$ (vertical) axes are labeled in cm.              }
         \label{hama}
   \end{figure}

%                                                One column figure
   \begin{figure}
   \centering
   \includegraphics[width=9cm]{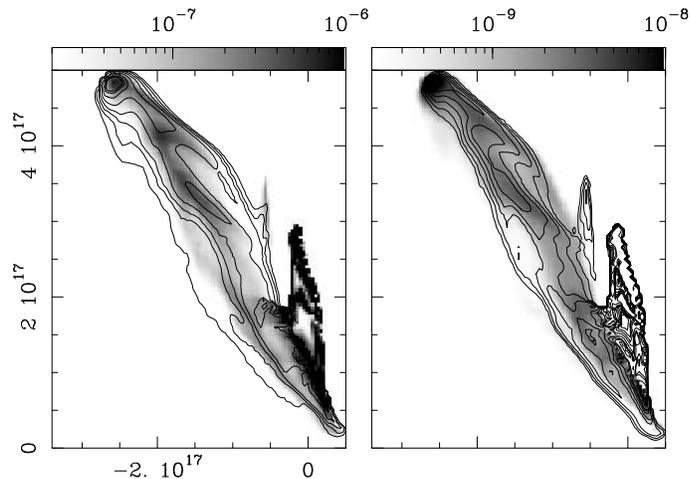}
   \caption{H$\alpha$ and H$_2$ 1-0 s(1) intensity maps
computed from model A for a $t=2600$~yr integration time (see
\S 3.2). The left plot shows the H$\alpha$ map in factor
of 2 contours and the H$_2$ map in the greyscale described
by the bar at the top of the plot. The right plot shows
the H$_2$ map in factor of 2 contours and the H$\alpha$
map in gresycale (corresponding to the bar at the top of the plot).
The greyscales (given in erg~cm$^{-2}$~s$^{-1}$~sterad$^{-1}$ by the
corresponding bars) and the contours of the intensity maps of a given
line correspond to the same range of intensities.
The $x$ (horizontal) and $z$ (vertical) axes are labeled in cm.}
              \label{h2ma}
    \end{figure}
For the H$\alpha$ emission line coefficient, we have included the
recombination cascade and the $n=1\to 3$ collisional excitation
(using the corresponding excitation coefficient of Giovanardi
\& Palla 1989). The H$_2$ 1-0 s(1) coefficient was computed by
solving the level population equations according to the prescription
of Draine et al. (1983, corrected according to Flower et al. 1986).
This is by no means the more accurate calculation of H$_2$ level
populations available (see, e.~g., Flower \& Pineau des For\^ets
1999), but it is appropriate given the limited accuracy of our rather
low resolution numerical simulations.

We have then computed intensity maps by integrating the emission
coefficients along lines of sight. We have assumed
that the $y=0$ plane (which includes the axis of the incident flow
and the centre of the cloud) coincides with the plane of the sky.
This is probably a reasonable approximation to the orientation
of the HH~270/110 flow, since it is known that both HH~270 and
HH~110 approximately lie on the plane of the sky (Reipurth
et al. 1986).

\subsection{Intensity maps predicted from model A}

In Figure 6, we show the H$\alpha$ emission line maps predicted from
model A (see \S 2.2). From the bottom plot, it is completely clear
that the emission is dominated by the region in which the incident
jet beam is digging into the cloud. As the optical emission maps
of HH~110 do not show this emitting region, we have to conclude
that it has to be absorbed by the dust in the dense cloud (at least,
if we believe in the jet/cloud interaction model for this object).

In order to illustrate the effect on the H$\alpha$ maps of
such an extinction, we have computed emission maps including this
effect (top two maps of figure 6). We have computed the maps
assuming that the dust extinction coefficient is $\kappa_d=
10^{-20} (n_H/{\rm cm^{-3}})$~cm$^{-1}$, giving a $\tau_d=40$
optical depth through the radius of the cloud.

We should note that if we use the standard, $\kappa_d=
10^{-21} (n_H/{\rm cm^{-3}})$~cm$^{-1}$ dust absorption
coefficient, our cloud would have a $\tau_d=4$ central
optical depth. The region on the edge of the cloud
into which the jet is penetrating would then have a low optical
depth, producing little extinction of the emission from the
impact region. This, however, is not a major problem given the
fact that the dense cloud present in the HH~270/110 region is
by no means either spherical or homogeneous, and could easily
produce a large extinction towards the current jet/cloud impact region
(which would lie within the cloud).

The H$\alpha$ maps obtained from model A (for integration times
$t=2000$ and 2600~yr) computed with the dust extinction as
described above are shown in the two top frames of Figure 6. It is
clear that these maps do present a qualitative similarity to HH~110
(see, e.~g., Reipurth et al. 1996).

In agreement with the observations, the emitting region starts
with a bright rim (in contact with the surface of the cloud,
see figure 6) which
points to a broader structure (with a complex structure of
curved ridges) at larger distances from the impact region.
Also, there is a faint emission halo extending parallel to
the main emission structure on the side directed away from
the dense cloud. This is also in qualitative agreement
with the H$\alpha$ maps of HH~110 (Reipurth et al. 1996).

From the two time frames shown in Figure 6, we have computed
proper motions for the three main intensity maxima seen in the maps.
We have also computed proper motions for some of the local
maxima in the region in which the emission is in contact
with the cloud surface, but the resulting velocities lie between
3 and 8~km~s$^{-1}$, and have not been plotted in Figure 6.

The proper motions of the knots farther away from the cloud
(shown in figure 6) have values of 15 to 45~km~s$^{-1}$.
These proper motions are substantially lower than the
ones measured for the HH~110 knots, as Reipurth et al. (1996)
have found values ranging from 35 to 150 km~s$^{-1}$.

In Figure 7, we show a comparison between the H$\alpha$ and
the H$_2$ 1-0 s(1) intensity maps obtained for $t=2600$~yr. For
computing the H$_2$ map, we have considered an extinction equal
to $1/10$ of the visual extinction. This results in only small optical
depths towards the current jet/cloud impact region, so that the
emission from this region is clearly visible.

The H$_2$ emission in the base of the deflected jet region is much more
concentrated towards the surface of the cloud than the H$\alpha$ emission.
This result is in clear qualitative agreement with the morphology
observed in HH~110  (Davis et al. 1994; Noriega-Crespo et al. 1996).
At larger distances along the deflected jet flow, the H$_2$ and
H$\alpha$ emission show spatially coincident condensations, again
in agreement with the observations of HH~110.

%                                     Two column figure
   \begin{figure*}
   \centering
   \includegraphics[width=9cm,angle=-90]{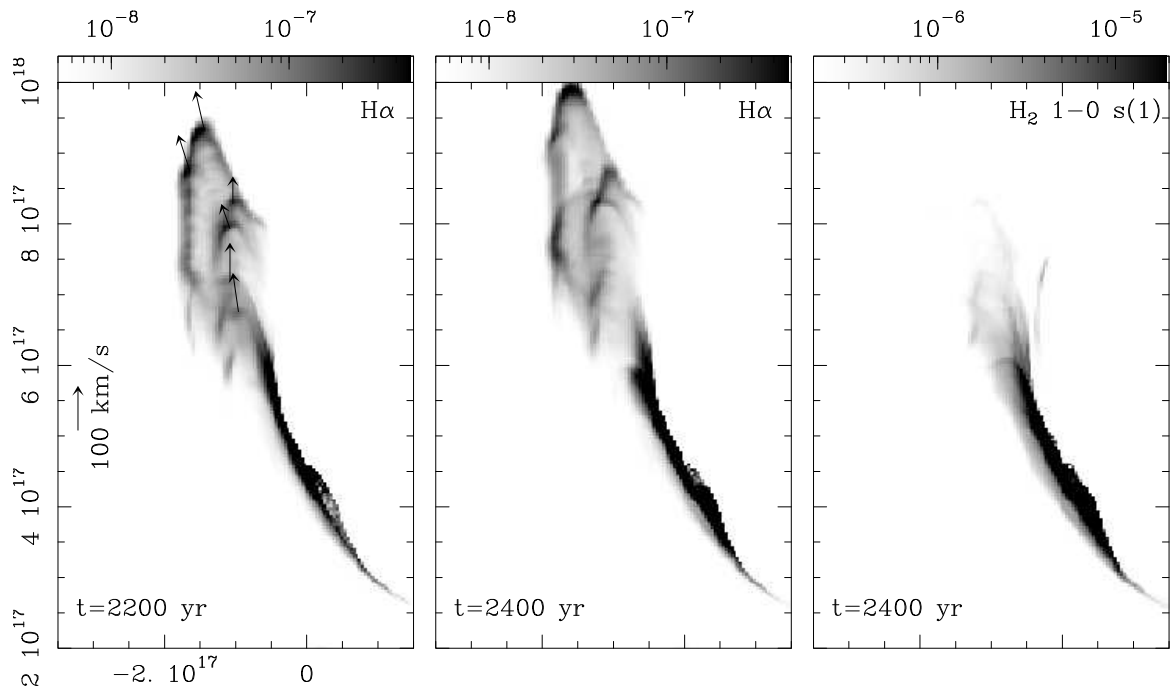}
   \caption{Intensity maps obtained from model B (see \S 3.3). The
left and central frames show the H$\alpha$ maps obtained for two
integration times (given on the bottom left of each frame). The right
hand side frame shows the H$_2$ 1-0 s(1) map obtained for $t=2400$~yr.
The maps are depicted with the greyscales given
(in erg~cm$^{-2}$~s$^{-1}$~sterad$^{-1}$) by the bars on the top of
each frame. The left plot shows the proper motions of several intensity
maxima, computed from the $t=2200$ and 2400~yr H$\alpha$ intensity maps.
The $x$ (horizontal) and $z$ (vertical) axes are labeled in cm.}
              \label{hah2m2}
    \end{figure*}

\subsection{Intensity maps predicted from model B}

In Figure~8, we show the H$\alpha$ intensity maps computed from
model B for $t=2200$ and 2400~yr integration times, and the
H$_2$ 1-0 s(1) map obtained for $t=2400$~yr. Because of the fact
that the jet/cloud impact region still lies on the surface of the cloud,
the effect of the extinction due to the dust present in the cloud
is not important, and has not been included. However, the
extinction would be important for maps computed for different orientations
of the jet/cloud structure with respect to the plane of the sky.

As a result of the
precession and ejection velocity time variability of the incident
jet (see \S 2.3), the emission maps show more complex structures
than the ones obtained from model A. In particular, one can clearly
see the emission from bow shocks around dense ``bullets'',
which result from the successive internal working
surfaces present in the incident jet. Even though the morphology
observed in the emission line maps of HH~110 is very complex (see,
e.~g., Reipurth et al. 1996), it does not appear to have such
bow shock structures. Actually, the intensity maps predicted from
model A do resemble the structure of HH~110 in a more convincing way.

An interesting feature of model B is that the proper motions
of the different intensity maxima (obtained by comparing
the $t=2200$ and 2400 yr H$\alpha$ intensity maps, see figure 8)
have values of close to 100~km~s$^{-1}$. These velocities are
in better agreement with the ones measured for HH~110 than
the ones obtained from model A (see \S 3.2 and Reipurth et al.
1996).

\section{Conclusions}

We have presented two jet/cloud collision 3D gasdynamic simulations~:
one with an incident jet with time-independent injection conditions
(model A), and a second one with a variable velocity, precessing
incident jet (model B). A $\rho_c/\rho_j=100$ cloud to initial
jet density ratio has been chosen for both models.

Model A produces a deflection of the jet beam only for a
$\sim 500$-1000~yr timescale, after which the incident jet starts
digging a straight tunnel through the dense cloud. At later times,
the deflected jet material continues to travel away from the
impact region, leaving behind a complex ``wake''.

Model B produces a broader jet/cloud impact region as a result of
the jet precession. This effect results in a longer timescale for
the duration of the jet deflection on the cloud surface (in fact,
the jet deflection is still occuring at the end of our
$t=3000$~yr numerical simulation).

Model B is more successful at reproducing the proper motions of
HH~110, giving $\sim 100$~km~s$^{-1}$ velocities for the H$\alpha$
intensity maxima along the deflected jet beam. Model A gives
velocities of $\sim 15$-45~km~s$^{-1}$, which are substantially lower
than the proper motion velocities of HH~110 (see Reipurth et al.
1996).

This difference between model A and model B is due to two effects.
The first effect is that in model A, the jet is
deflected only for a $\sim 500$-1000~yr timescale, and that
this deflected jet material then slows down as it interacts with
surrounding, environmental gas. In model B, successive deflected
``bullets'' (i.~e., internal working surfaces)
travel into the low density region left behind
by the passage of the head of the deflected jet, and do not
interact directly with the higher density environment. The
second effect is that because of the precession of model B,
some of the bullets have trajectories which are more tangential
to the surface of the molecular cloud. These more tangential
bullets are less deflected, and therefore preserve larger
velocities than the ones that have a more normal incidence
on the cloud surface (or than the deflected jet of model A).

However, in most other counts, model A is more successful than
model B at reproducing the observations of HH~110~:
\begin{itemize}
\item the general qualitative appearance of the H$\alpha$ maps
is in better agreement with the HH~110 H$\alpha$ images,
\item the features of the H$_2$ 1-0 s(1) emission and their
spatial relation to the H$\alpha$ emission also are in good
qualitative agreement with HH~110,
\item the maps that better resemble HH~110 correspond to times
(e.~g., the $t=2400$~yr frame of figures 1 and 6) in which the
impact region is already immersed within the cloud.
\end{itemize}
This last feature offers an interesting way of reconciling the
jet/cloud collision model with the HCO$^+$ observations of
Choi (2001).

In these observations, no HCO$^+$ emission was detected in the
region in which the axis of the ``incident'' HH~270 jet crosses
the axis of the ``deflected'' HH~110. Choi (2001) noted that this
appeared to be in disagreement with a jet/cloud collision model
for this system, as the cloud shock produced in the impact region
should indeed produce HCO$^+$ emission.

The situation found in model A, however, could indeed be in agreement
with the observations of Choi (2001). In this model, the impact
region does not lie in the point in which the incident and deflected
jet axes cross, but is instead located further along the axis
of the incident jet. Interestingly, Choi (2001) does find
substantial HCO$^+$ emission West of HH~110, approximately
aligned with HH~270.

As we have discussed in \S 3.2, the fact that the impact region
is not observed optically can in principle be a result of the
dust extinction in the dense cloud. Interestingly, some H$_2$ 1-0 s(1)
emission is apparently detected to the West of HH~110 (more or
less aligned with HH~270, see Noriega-Crespo et al. 1996), which
in principle might be associated with the impact region.

To conclude, we note the two main features of our results~:
\begin{itemize}
\item we find that the jet/cloud interaction model does reproduce
the H$\alpha$ and H$_2$ 1-0 s(1) emission observed in HH~110
in a qualitatively successful way,
\item our models show that the lack of a detected impact zone
in the incident/deflected jet crossing regions (Choi 2001) is
not a major problem for a jet/cloud collision model
(as this region could presently be displaced further into
the cloud).
\end{itemize}

Clearly, important questions remain about the more technical
aspects of our simulations. In particular, the H$_2$ emission from
the deflected jet comes from molecular cloud material which
has been entrained into the jet flow (see figures 2 and 5).
Even though we find that our models produce H$_2$ emission
structures in agreement with the observations of HH~110, the
accuracy of our rather low resolution simulations in reproducing
the entrainment process (which gives rise to this emission) is
somewhat questionable. Because of this, the present results have
to be taken with some caution.

\begin{acknowledgements}
The work of AR and PV was supported by CONACyT grants 34566-E
and 36572-E. AR acknowledges support from a fellowship of the
John Simon Guggenheim Memorial Foundation. PDM is a CONICET fellow.
The work of EMGDP has been partially supported by a grant
of the Brazilian Foundation FAPESP (13084-3).
The research of ANC was partially supported by NASA-APD Grant
NRA0001-ADP-096 and by the Jet Propulsion Laboratory, Caltech.
We thank Israel D\'\i az for installing the new computer with
which the numerical simulations were carried out.

\end{acknowledgements}

\appendix
\section{The atomic/ionic cooling rates}

For H~I and II, C~II and III and O~I and II,
we take the cooling rates (per atom and per electron) tabulated
by Raga et al. (1997). In order to
simplify the calculation of the cooling rate, we note that the
cooling per atom or ion starts to deviate from the low density
limit only for electron densities $n_e>10^4$~cm$^{-3}$. As such
electron densities are generally not found in HH objects, one
can safely consider only the low density limit.

In this limit, the cooling per atom and per electron is only a function
of the temperature, and one can then make simple parametric fits
for the cooling due to each of the considered species. The following
fits (giving the cooling rate in erg s$^{-1}$ cm$^3$)
have errors smaller than 10\%\ over the full $10^3\to 10^6$~K range
(unless otherwise noted)~:

\vskip.5cm
\noindent 1. Collisional excitation of HI:

\begin{equation}
\log_{10}\left({L_{HI}\over n_e n_{HI}}\right)=-50+32.3(1-t)-
1180(1-t^{0.0001}),
\end{equation}
with $t=1590$K$/T$,

\vskip.5cm
\noindent 2. Collisional ionisation of HI:

\begin{equation}
L_{ion,HI}=n_e n_{HI} q(T)\chi_H\,,
\end{equation}
where $q(T)$ is the collisional ionisation coefficient and $\chi_H$
the ionisation potential of H.

\vskip.5cm
\noindent 3. Radiative recombination of HII:

We have included the classical interpolation formula of
Seaton (1959).

\vskip.5cm
\noindent 4. Collisional excitation of O I (electrons):

$L=L_1+L_2$ with
\begin{equation}
\log_{10}\left({L_1\over n_e n_{OI}}\right)=-23.95+1.23t_1+0.5{t_1}^{10}\,,
\end{equation}
\begin{equation}
\log_{10}\left({L_2\over n_e n_{OI}}\right)=-21.05+1.2t_2+1.2\,
[\max(t_2,0)]^5\,,
\end{equation}
with $t_1=1-100\,{\rm K}/T$ and $t_2=1-10^4\,{\rm K}/T$.

\vskip.5cm
\noindent 5. Collisional excitation of O I (neutrals)

\begin{equation}
\log_{10}\left({L\over n_{HI} n_{OI}}\right)=10.3t+t^8-34.4\,,
\end{equation}
with $t=10\,{\rm K}/T$. This interpolation formula fits the computed
cooling in the $10\to 10^5$~K temperature range.

\vskip.5cm
\noindent 6. Collisional excitation of O II

$L=L_1+L_2$ with
\begin{equation}
\log_{10}\left({L_1\over n_e n_{OII}}\right)=7.9t_1-26.8\,,
\end{equation}
\begin{equation}
\log_{10}\left({L_2\over n_e n_{OII}}\right)=1.9\,{t_2\over
|t_2|^{0.5}}-20.5\,,
\end{equation}
with $t_1=1-2000\,{\rm K}/T$ and $t_2=1-{5\times 10^4}\,{\rm K}/T$.

\vskip.5cm
\noindent 7. Collisional excitation of C II

$$\log_{10}\left({L_1\over n_e n_{CII}}\right)=$$
\begin{equation}
-23.65+1.2\,[\max(t-2),0]^{[1.5-0.25\max(t-4,0)]}\,,
\end{equation}
with $t=\log_{10} T$~[K].

\vskip.5cm
\noindent 8. Collisional excitation of C III

\begin{equation}
\log_{10}\left({L_1\over n_e n_{CIII}}\right)=
-20.8+3.9(t-4)-1.37(t-4)^2\,,
\end{equation}
with $t=\min(\log_{10} T\,{\rm [K]},5.4)$.

\vskip.5cm
\noindent 9. Parametrized cooling for OIII

We have replaced the real cooling due to collisional excitation
of OIII with a function that resembles the coronal ionisation
equilibrium for $T>10^5$~K. In this way, the computed cooling has
a transition to the coronal equilibrium cooling when all O becomes
OIII and all C becomes CIV (with which no cooling is computed).
The adopted functional form is $L=L_1+L_2$ with
\begin{equation}
\log_{10}\left({L_1\over n_e n\,y_{OIII}}\right)=-21.4-3.5(t-5.4)^2\,,
\end{equation}
\begin{equation}
\log_{10}\left({L_2\over n_e n\,y_{OIII}}\right)=-21.7-0.7(t-6.5)^2\,,
\end{equation}
where $t=\log_{10} T$~[K] and $y_{OIII}$ is the OIII ionisation
fraction.


\begin{thebibliography}{}

\bibitem[1973]{Aldrovandi73} Aldrovandi, S. M. V., P\'equignot, D.,
1973, A\&A, 25, 137

\bibitem[1976]{Aldrovandi76} Aldrovandi, S. M. V., P\'equignot, D.,
1976, A\&A, 47, 321

\bibitem[1996]{Canto96} Cant\'o, J., Raga, A. C., 1996,
MNRAS, 280, 559

\bibitem[2001]{Choi01} Choi, M., 2001, ApJ, 550, 817

\bibitem[1970]{Cox70} Cox, D., 1970, Ph.D. Thesis (Univ. of California)

\bibitem[1994]{Davis94} Davis, C. J., Mundt, R., Eisl\"offel, J.,
1994, ApJ, 437, L55

\bibitem[1983]{Draine83} Draine, B. T., Roberge, W. G.,
Dalgarno, A., 1983, ApJ, 264, 485

\bibitem[1986]{Flower86} Flower, D. R., Pineau des For\^ets,
G., Hartquist, T. W., 1986, MNRAS, 218, 729

\bibitem[1999]{Flower99} Flower, D. R., Pineau des For\^ets,
G., 1999, MNRAS, 308, 271

\bibitem[1989]{Giovanardi89} Giovanardi, C., Palla, F., 1989,
A\&AS, 77, 157

\bibitem[1999]{Gouveia99} de Gouveia Dal Pino, E. M., 1999,
ApJ, 526, 862

\bibitem[1999]{Hurka99} Hurka, J. D., Schmid-Burgk, J., Hardee,
P. E., 1999, A\&A, 343, 558

\bibitem[1996]{Noriega96} Noriega-Crespo, A., Garnavich, P. M.,
Raga, A. C., Cant\'o, J., B\"ohm, K. H., 1996, ApJ, 462, 804

\bibitem[1995]{Raga95} Raga, A. C., Cant\'o, J., 1995,
RMxAA, 31, 51

\bibitem[1995]{Raga95b} Raga, A. C., Taylor, S. D., Cabrit, S.,
Biro, S., 1995, A\&A, 833, 843.

\bibitem[1996]{Raga96} Raga, A. C., Cant\'o, J., 1996,
MNRAS, 280, 567

\bibitem[1997]{Raga97} Raga, A. C., Mellema, G., Lundqvist, P.,
1997, ApJS, 109, 517

\bibitem[2000]{Raga00} Raga, A. C., Navarro-Gonz\'alez, R.,
Villagr\'an-Muniz, M., 2000, RMxAA, 36, 67

\bibitem[2000]{Raga01} Raga, A. C., Sobral, H.,
Villagr\'an-Muniz, M., Navarro-Gonz\'alez, R., Masciadri, E., 2001,
MNRAS, 324, 206

\bibitem[1986]{Reipurth86} Reipurth, B., Bally, J., 1986,
Nature, 320, 336

\bibitem[1996]{Reipurth96} Reipurth, B., Raga, A. C., Heathcote, S.,
1996, AJ, 311, 989

\bibitem[1998]{Rod98} Rodr\'\i guez, L. F., Reipurth, B., Raga, A. C.,
Cant\'o, J., 1998, RMxAA, 34, 69

\bibitem[1959]{Seaton59} Seaton, M. J., 1959, MNRAS, 119, 81

\bibitem[2000]{Sobral00} Sobral, H., Villagr\'an-Muniz, M.,
Navarro-Gonz\'alez, R., Raga, A. C., 2000, App. Phys. Lett., 77, 3158

\end{thebibliography}
\end{document}